\documentclass[journal,comsoc,onecolumn,twoside,10pt]{IEEEtran}
\usepackage{graphicx}
\usepackage{caption}
\usepackage{epstopdf}
\usepackage{amsmath}
\usepackage{mathtools,amssymb}
\usepackage{romannum}
\begin{document}
\title{\LARGE Study of Iterative Detection and Decoding for Large-Scale MIMO Systems with 1-Bit ADCs}

\author{Zhichao~Shao,~Rodrigo~C.~de~Lamare,~\IEEEmembership{Senior~Member,~IEEE}~and~Lukas~T.~N.~Landau,~\IEEEmembership{Member,~IEEE}\vspace{-0.5cm}
    \thanks{The authors are with the Pontifical Catholic University of Rio de Janeiro, Centre for Telecommunications Studies, Rio de Janeiro, CEP 22453-900, Brazil (e-mail: {zhichao.shao;delamare;lukas.landau}@cetuc.puc-rio.br).}}

\maketitle

\begin{abstract}
We present a novel iterative detection and decoding scheme for
uplink large-scale multiuser multiple-antenna systems. In order to
reduce the receiver's energy consumption and computational
complexity, 1-bit analog-to-digital converters are used in the
front-end. The performance loss due to the 1-bit quantization can be
mitigated by using large-scale antenna arrays. We propose a linear
low-resolution-aware minimum mean square error detector for soft
multiuser interference mitigation. Moreover, short block length
low-density parity-check codes are considered for avoiding high
latency. In the channel decoder, a quasi-uniform quantizer with
scaling factors is devised to lower the error floor of LDPC codes.
Simulations show good performance of the system in terms of bit
error rate as compared to prior work.
\end{abstract}

\begin{IEEEkeywords}
Large-scale multiple-antenna systems, 1-bit quantization, IDD schemes, MMSE detectors.\vspace{-0.3cm}
\end{IEEEkeywords}

\section{Introduction}

Large-scale multiple-antenna systems have been identified as a
promising technology for the next generation communication systems
\cite{Larrson,mmimo,wence15}. In fact, large spatial degrees of
freedom (DoFs) can increase the spectral and energy efficiency.
However, as the antennas scale up, the receiver design will become
more complex and the energy consumption will be higher. For
overcoming these issues, one solution is to use low-resolution
analog-to-digital converters (ADCs) at the receiver. As one extreme
case, 1-bit quantization can drastically simplify the receiver
design. Prior work on multiple-antenna systems with low-resolution
ADCs includes the studies in \cite{Landau17} and \cite{Nossek},
where in the latter the authors reported a novel linear minimum mean
square error (MMSE) precoder design for multiple-antenna systems
using 1-bit DAC/ADC both at the transmitter and the receiver. An
analytical approach to calculate a lower bound on capacity for a
wideband system with multiple-antenna and 1-bit ADCs, which employs
low-complexity channel estimation and symbol detection, has been
described in \cite{Mollen}.

In recent years, low-density parity-check (LDPC) codes have beenused
in many industry standards including DVB-S2 and IEEE 802.11n
(Wi-Fi). They have also been adopted for the next generation
communication systems, since they approach the Shannon capacity and
have low complexity. Compared to LDPC codes with large block length,
short block LDPC codes result in much lower latency. As one branch
of LDPC codes, regular LDPC codes, they have high error floor
phenomenon, which is commonly attributed to the existence of certain
error-prone structures in the corresponding Tanner graph. This is
partially because of trapping sets \cite{Richardson,memd} and
absorbing sets \cite{Dolecek}. The authors in \cite{Zhang} have
proposed a new LDPC decoder with low error floors and low
computational complexity. This approach quasi-uniformly quantizes
the passing messages into different ranges of reliability. It
extends the saturation level to prevent the messages from being
trapped and can be helpful for 1-bit quantized data. In receiver
designs with channel codes, it is often useful to employ iterative
detection and decoding (IDD) schemes
\cite{Wang,spa,mfsic,mbdf,dai15,qin16,Uchoa}. The key mechanism of
the IDD process is the soft information exchange between the
detector and the channel decoder, which leads to successive
performance improvement \cite{Wang}. The soft information exchanged
often has the form of log likelihood ratio (LLR) of a certain bit.
In \cite{Uchoa} an IDD algorithm with LDPC codes for
multiple-antenna systems under block-fading channels is proposed.
Results show that by properly manipulating the LLR output of decoder
the system performance can be largely improved.

In this work, we develop an IDD scheme for 1-bit quantized systems and derive a linear low-resolution-aware MMSE (LRA-MMSE) receive filter suitable for 1-bit ADCs and soft interference mitigation. Moreover, we also develop an adaptive decoding approach that combines a quasi-uniform quantization of the passing messages with adjustable scaling factors, which can avoid trapping sets and refine the exchange of LLRs between the detector and the decoder.

The rest of this paper is organized as follows: Section \Romannum{2} shows the system model and presents some statistical properties about 1-bit quantization. Section \Romannum{3} describes the derivation of the proposed LRA-MMSE detector and decoding technique. Section \Romannum{4} discusses the simulation results and section \Romannum{5} concludes the work.

Notation: Bold capital letters indicate matrices while vectors are in bold lowercase. $\mathbf{I}_n$ denotes a $n\times n$ identity matrix while $\mathbf{0}_n$ is a $n\times 1$ vector with all zeros. Additionally, diag$(\mathbf{A})$ is a diagonal matrix only containing the diagonal elements of $\mathbf{A}$.
\vspace{-0.15cm}
\begin{figure*}[!t]
    \centering
    \includegraphics[]{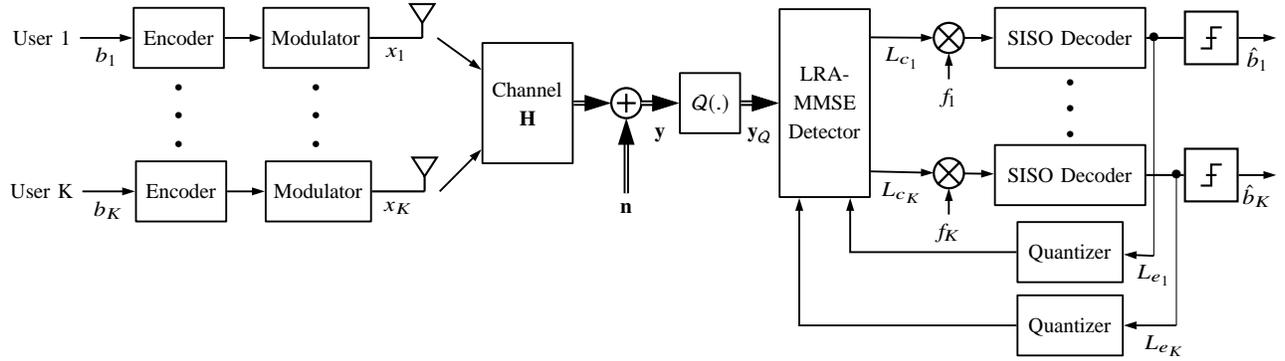}
    \caption{System model of a multi-user multiple-antenna system}
    \label{fig:transmitter}
    \vspace{-0.2cm}
\end{figure*}

\section{System Model and Statistical Properties of 1-bit Quantization}

A single-cell multi-user large-scale multiple-antenna uplink
scenario is considered, which is depicted in Fig. 1. There are $K$
single-antenna users and the receiver is equipped with $M$ antennas,
where $M\gg K$. The information symbols $b_k$ are firstly encoded by
each user's channel encoder and modulated to $x_k$ according to a
given modulation scheme. The transmit symbols $x_k$ have zero-mean
and the same energy $E[|x_k|^2]=\sigma_x^2$. The modulated symbols
are then transmitted over block-fading channels. The vector
$\mathbf{n}\sim \mathcal{CN}(\mathbf{0}_M,\sigma^2_n\mathbf{I}_M)$
contains independent and identically distributed (IID) complex
Gaussian random variables with zero mean and variance $\sigma^2_n$.
The received unquantized signal is given by
\begin{equation}
    \mathbf{y}=\mathbf{H}\mathbf{x}+\mathbf{n}=\sum_{k=1}^{K}\mathbf{h}_kx_k+\mathbf{n},
\end{equation}
where $\mathbf{H}\in \mathbb{C}^{M\times K}$ is the channel matrix. $\mathcal{Q}(.)$ represents the 1-bit quantization. The real and imaginary parts of the unquantized signal $\mathbf{y}$ are element-wisely quantized to $\{\pm\frac{1}{\sqrt{2}}\}$ based on a threshold. The resulting quantized signal $\mathbf{y}_\mathcal{Q}$ is
\begin{equation}
\mathbf{y}_\mathcal{Q}=\mathcal{Q}\left(\mathfrak{R}\{\mathbf{y}\}\right) + j\mathcal{Q}\left(\mathfrak{I}\{\mathbf{y}\}\right),
\label{system_model_quantize}
\end{equation}
where $\mathfrak{R}\{\cdot\}$ and $\mathfrak{I}\{\cdot\}$ get the real and imaginary part, respectively.

Since quantization strongly changes the properties of signals, we show here some statistical properties of quantization for a Gaussian input signal. For 1-bit quantization and Gaussian inputs, the cross-correlation between the unquantized signal $\mathbf{s}$ with covariance matrix $\mathbf{C}_\mathbf{s}$ and its 1-bit quantized signal $\mathbf{s}_\mathcal{Q}$ is described by \cite{Bussgang}
\begin{equation}
    \mathbf{C}_{\mathbf{s}_\mathcal{Q}\mathbf{s}}=\sqrt{\frac{2}{\pi}}\mathbf{K}\mathbf{C}_{\mathbf{s}},\mbox{where } \mathbf{K}=\text{diag}(\mathbf{C}_{\mathbf{s}})^{-\frac{1}{2}}.
\end{equation}
Furthermore, the covariance matrix of the 1-bit quantized signal $\mathbf{s}_\mathcal{Q}$ is given by \cite{Jacovitti}
\begin{equation}
\mathbf{C}_{\mathbf{s}_\mathcal{Q}}=\frac{2}{\pi}\left(\text{sin}^{-1}\left(\mathbf{K}\mathfrak{R}\{\mathbf{C}_{\mathbf{s}}\}\mathbf{K}\right)+j\text{sin}^{-1}\left(\mathbf{K}\mathfrak{I}\{\mathbf{C}_{\mathbf{s}}\}\mathbf{K}\right)\right).
\end{equation}
\section{Proposed Iterative Detection and Decoding}
\subsection{Proposed LRA-MMSE Detector}
Inspired by prior work on IDD schemes \cite{Wang} \cite{Uchoa}, we propose a LRA-MMSE detector, which employs a modified linear MMSE receive filter and performs soft parallel interference cancellation. The soft estimate of the $k$-th transmitted symbol is firstly calculated based on the extrinsic LLR $L_{e_k}$ provided by the channel decoder from a previous stage:
\begin{equation*}
\tilde{x}_k=\sum_{x\in \mathcal{A}}x\text{Pr}(x_k=x)=\sum_{x\in \mathcal{A}}x\left(\prod_{l=1}^{M_c}\left[1+\text{exp}(-x^lL_{e_k}^{l})\right]^{-1}\right),
\end{equation*}
where $\mathcal{A}$ is the complex constellation set with $2^{M_c}$ possible points. The symbol $x^l$ corresponds to the value $(+1,-1)$ of the $l$th bit of symbol $x$. Denote $\mathbf{\tilde{x}} = [\tilde{x}_1,...,\tilde{x}_K]^T$ and
\begin{equation}
\mathbf{\tilde{x}}_k = \mathbf{\tilde{x}} - \tilde{x}_k\mathbf{e}_k,
\end{equation}
where $\mathbf{e}_k$ is a column vector with all zeros, except that the $k$th element is equal to 1. For each user $k$, the interference from the other $K-1$ users is canceled according to
\begin{equation}
\mathbf{y}_{\mathcal{Q}_k}=\mathbf{y}_\mathcal{Q}-\sum_{j=1,j\neq k}^{K}\tilde{x}_j\mathbf{h}_j=\mathbf{y}_\mathcal{Q}-\mathbf{H}\mathbf{\tilde{x}}_k.
\end{equation}
Note that when no prior information is given, $\mathbf{y}_{\mathcal{Q}_k}=\mathbf{y}_{\mathcal{Q}}$. The linear LRA-MMSE filter is then applied to $\mathbf{y}_{\mathcal{Q}_k}$, to obtain
\begin{equation}
\hat{x}_k=\mathbf{w}_k^H\mathbf{y}_{\mathcal{Q}_k},
\label{estimated_symbol}
\end{equation}
where $\mathbf{w}_k$ is chosen to minimize the mean square error (MSE) between the transmitted symbol $x_k$ and the filter output, i.e.
\begin{equation}
\mathbf{w}_k=\arg\min_{\mathbf{w'}_k} E\left[\left\vert\left\vert x_k-\mathbf{w'}_k^{H}\mathbf{y}_{\mathcal{Q}_k}\right\vert\right\vert^2\right].
\end{equation}
The solution of the LRA-MMSE receive filter is given by
\begin{equation}
\mathbf{w}_k=\mathbf{C}_{\mathbf{y}_{\mathcal{Q}_k}}^{-1}\mathbf{c}_{x_k\mathbf{y}_{\mathcal{Q}_k}},
\end{equation}
where the covariance matrix is
\begin{equation}
\mathbf{C}_{\mathbf{y}_{\mathcal{Q}_k}}=\mathbf{C}_{\mathbf{y}_\mathcal{Q}}-\left(\mathbf{C}_{\mathbf{y}_\mathcal{Q}\mathbf{\tilde{x}}_k}\mathbf{H}^H\right)^H-\mathbf{C}_{\mathbf{y}_\mathcal{Q}\mathbf{\tilde{x}}_k}\mathbf{H}^H+\mathbf{H}\mathbf{C}_{\mathbf{\tilde{x}}_k}\mathbf{H}^H
\label{covariance_in_lrammse}
\end{equation}
and the cross-correlation vector is
\begin{equation}
\mathbf{c}_{x_k\mathbf{y}_{\mathcal{Q}_k}}=\sigma_x^2\sqrt{\frac{2}{\pi}}\mathbf{K}\mathbf{h}_k,\quad \text{with}\quad\mathbf{K}=\text{diag}\left(\mathbf{C}_{\mathbf{y}}\right)^{-\frac{1}{2}}.
\end{equation}
In (\ref{covariance_in_lrammse}), the covariance matrix of the quantized data vector $\mathbf{y}_\mathcal{Q}$ is described by
\begin{equation}
\mathbf{C}_{\mathbf{y}_\mathcal{Q}}=\frac{2}{\pi}\left(\text{sin}^{-1}\left(\mathbf{K}\mathfrak{R}\{\mathbf{C}_{\mathbf{y}}\}\mathbf{K}\right)+j\text{sin}^{-1}\left(\mathbf{K}\mathfrak{I}\{\mathbf{C}_{\mathbf{y}}\}\mathbf{K}\right)\right),
\label{red}
\end{equation}
and the cross-correlation vector between $\mathbf{y}_\mathcal{Q}$ and $\mathbf{\tilde{x}}_k$ is
\begin{equation}
\mathbf{C}_{\mathbf{y}_\mathcal{Q}\mathbf{\tilde{x}}_k}=\sqrt{\frac{2}{\pi}}\mathbf{KH}\mathbf{C}_{\mathbf{\tilde{x}}_k}.
\label{red3}
\end{equation}
Note that $\mathbf{C}_{\mathbf{y}}$ is the covariance matrix of the unquantized data vector $\mathbf{y}$, which leads to the following result
\begin{equation}
\mathbf{C}_{\mathbf{y}}=E\left[\left(\mathbf{H}\mathbf{x}+\mathbf{n}\right)\left(\mathbf{H}\mathbf{x}+\mathbf{n}\right)^H\right]=\sigma_x^2\mathbf{HH}^H+\sigma_n^2\mathbf{I}_M.
\label{red6}
\end{equation}

In order to calculate $P(\hat{x}_k|x)$, we use the Cramer's central limit theorem \cite{cramer}: the LRA-MMSE filter output can be approximated by a complex Gaussian distribution due to the large number of independent variables. The mean and variance of the estimated symbol $\hat{x}_k$, which is conditioned on the transmitted symbol $x$, are given respectively by
\begin{equation}
\mu_k \overset{\Delta}{=} E\left[\hat{x}_k\vert x\right] =\mathbf{w}_k^H\left(\mathcal{Q}\left(\mathbf{h}_kx+\mathbf{H}\mathbf{\tilde{x}}_k\right)-\mathbf{H}\mathbf{\tilde{x}}_k\right)
\end{equation}
\begin{equation}
\eta_k^2 \overset{\Delta}{=} \text{var}\left[\hat{x}_k\vert x\right] = \mathbf{w}_k^H\mathbf{c}_{x_k\mathbf{y}_{\mathcal{Q}_k}}-\left(\mathbf{w}_k^H\mathbf{c}_{x_k\mathbf{y}_{\mathcal{Q}_k}}\right)^2.
\end{equation}
Therefore, the likelihood function can be approximated by
\begin{equation}
P(\hat{x}_k|x)\simeq\frac{1}{\pi\eta_k^2}\text{exp}\left(-\frac{1}{\eta_k^2}\left\vert\hat{x}_k-\mu_k\right\vert^2\right).
\end{equation}
Then the LLR computed by the LRA-MMSE detector for the $l$-th bit ($l\in \{1,...,M_c\}$) of the symbol $\hat{x}_k$ is given by
\begin{equation}
\begin{aligned}
L^{l}_{c_k}&=\log\frac{\text{Pr}\left(b^{l}_k=+1\vert\hat{x}_k\right)}{\text{Pr}\left(b^{l}_k=-1\vert\hat{x}_k\right)}-\log\frac{\text{Pr}\left(b^{l}_k=+1\right)}{\text{Pr}\left(b^{l}_k=-1\right)}\\&=\log\frac{\sum_{x\in \mathcal{A}^{+1}_l} P\left(\hat{x}_k|x\right)\text{Pr}\left(x\right)}{\sum_{x\in \mathcal{A}^{-1}_l} P\left(\hat{x}_k|x\right)\text{Pr}\left(x\right)}-L_{e_k}^{l},
\end{aligned}
\end{equation}
where $\mathcal{A}^{+1}_l$ is the set of hypotheses $x$ for which the $l$-th bit is +1 and $\mathcal{A}^{-1}_l$ is similarly defined.
\vspace{-0.45cm}

\subsection{Proposed Soft Information Processing and Decoding}
The soft information provided by the LRA-MMSE detector is then fed into a channel decoder that adaptively scales the input LLRs and quasi-uniformly quantizes the messages.

\subsubsection{Iterative Decoder}
The decoding method is based on message passing, which iteratively computes the distributions of variables in graph-based models. In the system we have used the box-plus sum product algorithm (SPA) \cite{Hu}, which is an approximation of SPA decoding. One drawback of SPA is the hyperbolic tangent function, which has numerical saturation problems when computed with finite precision. To avoid such problems, thresholds on the magnitudes of messages must be applied. In the box-plus SPA, the message sent from check node (CN) $j$ to variable node (VN) $i$ is
\begin{equation}
L_{j\rightarrow i} = \boxplus_{i'\in N(j)\backslash i} L_{i'\rightarrow j},
\vspace{-0.1cm}
\end{equation}
where $\boxplus$ is the pairwise "box-plus" operator defined as
\begin{equation}
\begin{aligned}
x\boxplus y&=\log\left(\frac{1+e^{x+y}}{e^x+e^y}\right)\\&=\text{sign}(x)\text{sign}(y)\min\left(\vert x\vert,\vert y\vert\right)\\&\quad+\log\left(1+e^{-\vert x+y\vert}\right)-\log\left(1+e^{-\vert x-y\vert}\right).
\end{aligned}
\vspace{-0.1cm}
\end{equation}
The message from VN $i$ to CN $j$ is then calculated as
\begin{equation}
L_{i\rightarrow j}=L_i + \sum_{j'\in N(i)\backslash j}L_{j'\rightarrow i},
\vspace{-0.1cm}
\end{equation}
where $L_i$ is the LLR at VN $i$. The quantity $j'\in N(i)\backslash j$ represents all CNs connected to VN $i$ except CN $j$.

\subsubsection{Quasi-uniform Quantizer}
This quantizer is used both in the decoder and the extrinsic message quantizer to refine or compensate for the effect of 1-bit quantization on the LLRs. The algorithm is based on the quasi-uniform quantization in \cite{Zhang}, which represents a compromise between conflicting objectives of retaining fine precision, allowing large dynamic range and implementation complexity. It is a combination of non-uniform and uniform quantization and realized as follows:
\begin{equation*}
Q^*_{\Delta}(L_{c_k}) =
\begin{cases}
d^{N+1}N\Delta       & \quad \text{if } d^{N+1}N\Delta\le L_{c_k}\\
d^rN\Delta  & \quad \text{if } d^{r}N\Delta\le L_{c_k}<d^{r+1}N\Delta, \\&\quad\text{for } N\ge r \ge1 \\
Q_{\Delta}(L_{c_k})  & \quad \text{if } -dN\Delta<L_{c_k}<dN\Delta\\
-d^rN\Delta  & \quad \text{if } -d^{r+1}N\Delta\le L_{c_k}<-d^{r}N\Delta,\\&\quad \text{for } 1\le r \le N \\
-d^{N+1}N\Delta       & \quad \text{if } L_{c_k} \le -d^{N+1}N\Delta\\
\end{cases}
\end{equation*}
with
\begin{equation*}
Q_{\Delta}(L_{c_k}) =
\begin{cases}
N\Delta       & \quad \text{if } N\Delta-\frac{\Delta}{2}\le L_{c_k}\\
m\Delta  & \quad \text{if } m\Delta-\frac{\Delta}{2}\le L_{c_k}<m\Delta+\frac{\Delta}{2}, \\&\quad\text{for } N> m >0 \\
0  & \quad \text{if } -\frac{\Delta}{2}<L_{c_k}<\frac{\Delta}{2}\\
-m\Delta  & \quad \text{if } m\Delta-\frac{\Delta}{2}< L_{c_k}\le m\Delta+\frac{\Delta}{2},\\&\quad \text{for } -N< m < 0 \\
-N\Delta       & \quad \text{if } L_{c_k} \le -N\Delta+\frac{\Delta}{2}\\
\end{cases}
\end{equation*}
where $d$ is the growth rate parameter, $\Delta$ is the step size, $N$ is the total number of bits for representing each range and $L_{c_k}$ is the passing message at the $k$-th decoder.

\subsubsection{Adaptive Scaling Factors}
For improving the decoding performance we have deployed two scaling factors, which are obtained offline and online, respectively.
\begin{itemize}
    \item Offline Scaling Factor: This factor is utilized to correct LLR values used in iterative decoding based on the LLR distribution \cite{Alvarado}. In the training phase, data packets are sent to the receiver for obtaining sufficient LLR statistics. For a given SNR, the following steps are carried out:
    \begin{itemize}
        \item[1)] Calculate the probabilities of $\text{Pr}(L_{c_k}\vert b_k)$ conditioned on transmitted bits $b_k$ through histograms.
        \item[2)] Obtain $f(L_{c_k})=\log\frac{\text{Pr}(L_{c_k}|b_k=1)}{\text{Pr}(L_{c_k}|b_k=0)}$.
        \item[3)] Employ the approximation $f(L_{c_k})=\alpha_k L_{c_k}$.
    \end{itemize}
    This factor $\alpha_k$ is only applied in the first iteration at the $k$th decoder input during the data transmission phase. Moreover, the scaled mean absolute value $\alpha_k\overline{L}_{c_k}$ for each user is stored for calculating the online scaling factor.
    \item Online Scaling Factor: The factor $f_k$ is calculated at the $k$th decoder input in the second iteration and applied for all the iterations except the first iteration. It aims to correct the LLR errors caused by quantizer. The scaled LLR should be approximated to the scaled LLR in 3). We propose a linear scaling factor that is calculated as:
    \begin{equation}
    f_k = \alpha_k\overline{L}_{c_k}/\overline{L}^{\text{2nd iteration}}_{c_k},
    \end{equation}
    where $\overline{L}^{\text{2nd iteration}}_{c_k}$ is the mean absolute value of LLRs for the $k$th user in the second iteration.
\end{itemize}

\section{Numerical Results}
We consider a short length regular LDPC code with block length $n = 512$ and rate 1/2. The modulation scheme is QPSK and the parameters of the quasi-uniform quantizer are $\Delta=0.25$, $d=1.3$ and $N=6$. The channel is assumed to experience block fading and is modeled by IID circularly symmetric complex Gaussian random variables with zero mean and unit variance. The channel matrix is estimated unless otherwise specified through the Bussgang-based LMMSE (BLMMSE) channel estimator \cite{Li}. During training, all $K$ users simultaneously transmit $\tau$ pilot symbols to the receiver. To match the matrix form to the vector form of (\ref{system_model_quantize}), the vectorized received signal is described by
\begin{equation*}
\mathbf{y}_{\mathcal{Q}_p}=\mathcal{Q}\left(\mathbf{y}_p\right)=\mathcal{Q}\left(\tilde{\mathbf{X}}_p\mathbf{h}+\mathbf{n}_p\right),
\end{equation*}
where $\tilde{\mathbf{X}}_p = (\mathbf{X}^T_p\otimes\mathbf{I}_M)\in \mathbb{C}^{M\tau\times KM}$ is the modified pilot matrix. The vector $\mathbf{h}\in \mathbb{C}^{KM\times 1}$ is the vectorized channel matrix $\mathbf{H}$. With the assumption $\mathbf{C}_{\mathbf{h}}=\mathbf{I}_{KM}$, the estimated channel vector is
\begin{equation}
    \hat{\mathbf{h}} = \left(\mathbf{A}_p\tilde{\mathbf{X}}_p\right)^H\mathbf{C}_{\mathbf{y}_{\mathcal{Q}_p}}^{-1}\mathbf{y}_{\mathcal{Q}_p},
\vspace{-0.2cm}
\end{equation}
where $\mathbf{A}_p = \sqrt{\frac{2}{\pi}}\text{diag}(\mathbf{C}_{\mathbf{y}_p})^{-\frac{1}{2}}$. $\mathbf{C}_{\mathbf{y}_{\mathcal{Q}_p}}$ and $\mathbf{C}_{\mathbf{y}_p}$ are calculated according to (\ref{red}) and (\ref{red6}), respectively.

\begin{figure}[!t]
    \centering
    \includegraphics[]{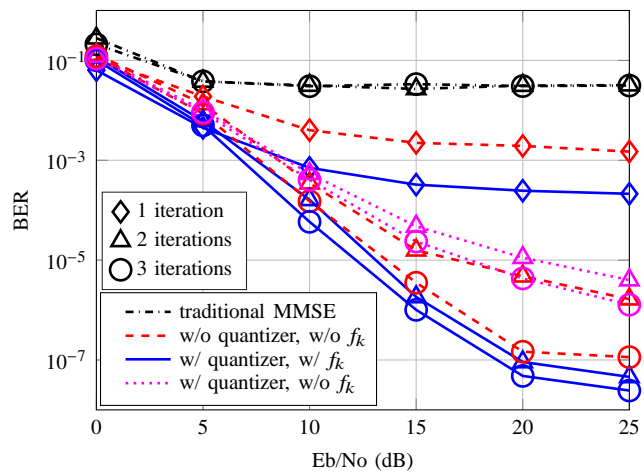}
    \vspace{-0.1cm}
    \caption{$K = 12$ and $M = 32$. BER performances of IDD schemes under the perfect CSI. (The single-user bound curve (no interference) is about 20dB to the left for the same BER.)}
    \label{fig:BER_IDD-with-quantization}
    \vspace{-0.4cm}
\end{figure}

\begin{figure}[!t]
    \centering
    \includegraphics[]{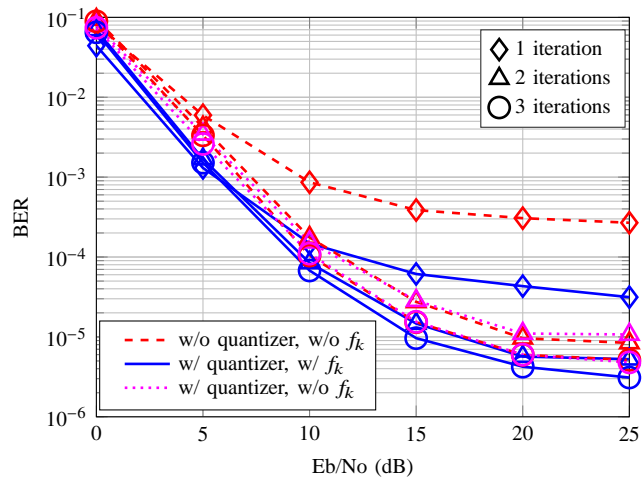}
    \vspace{-0.1cm}
    \caption{$K = 9$ and $M = 32$. BER performances of IDD schemes using BLMMSE channel estimation with $\tau = 70$ pilot symbols. (The single-user bound curve (no interference) is about 16dB to the left for the same BER.)}
    \label{fig:BER_IDD-with-quantization-imperfect-CSI}
    \vspace{-0.3cm}
\end{figure}

The BER performances of IDD schemes under perfect channel state information (CSI) are shown in Fig. \ref{fig:BER_IDD-with-quantization}. It can be seen that the proposed LRA-MMSE detector obtains a large gain compared to the traditional MMSE one. Recently, the authors in \cite{Kim} have proposed a near maximum likelihood (ML) detector with a 2dB performance gain, but the computational complexity ($\mathcal{O}(2^{KM_c}M^2)$) is much higher than that of the LRA-MMSE detector ($\mathcal{O}(M^3)$). Moreover, Fig. \ref{fig:BER_IDD-with-quantization} also depicts the BER performances of IDD schemes with and without quasi-uniform quantizer and scaling factors, which shows the system has a significant performance gain after 2 iterations. These results also demonstrate that the quantizer and the scaling factors offer extra performance gains. Fig. \ref{fig:BER_IDD-with-quantization-imperfect-CSI} illustrates the system performance using BLMMSE channel estimation, where $\tau = 70$ pilot symbols are used in each block.

\vspace{-0.1cm}
\section{Conclusion}
In this work, we have developed an IDD scheme for 1-bit quantized systems and proposed a LRA-MMSE detector for 1-bit systems. The simulation results have shown a great performance gain after several iterations. Moreover, we have devised an adaptive channel decoder using a quantizer together with scaling factors for further performance improvement.

\bibliographystyle{IEEEtran}
\bibliography{bib-refs-ZS}
\end{document}